# 基于知识"非常规重组"的学术论文新颖性测度指标研究

梁国强，孙健，林歌歌，张硕

（北京工业大学经济与管理学院，北京，100124）

**摘　要**　科学发展离不开对已有知识的重新组合，而新颖的研究通常依赖对现有知识的"非常规重组"。本研究提出知识离心度指标，通过测度待评价成果对已有知识体系的偏离程度，对待评价成果的新颖性进行及时评价。研究以 2005 年、2010 年、2015 年、2020 年和 2025 年发表在 Science 和 Nature 上的研究型论文、Top1%高被引论文和零被引论文作为样本数据，计算上述论文的知识离心度，并对其潜在的影响因素进行分析。结果显示，团队规模对论文新颖性产生负向影响（β=-0.005, p<0.001），即团队规模越大，越不利于提升成果的新颖性；参考文献数量与论文新颖性呈正向关系（β=0.027, p<0.001），当论文的参考文献数量越多时，论文的新颖性会在一定程度提升。本研究提出的知识离心度指标"及时性"和"可操作性"强，能够在论文发表时即可对其新颖性进行评价。

**关键词**　参考文献；知识离心度；非常规重组；团队规模

## Research on Novelty Measurement Indicator of Academic Papers Based on the Atypical Recombination of Knowledge

Liang Guoqiang, Sun Jian, Lin Gege, Zhang Shuo

(School of Economics and Management, Beijing University of Technology, Beijing 100124, China)

**Abstract:** The advancement of science is inherently dependent on the recombination of existing knowledge, and innovative research typically relies on the "atypical recombination" of established knowledge bases. This study introduces a Knowledge Eccentricity to enable timely assessment of the novelty of research outputs by quantifying their degree of deviation from the existing knowledge system. For empirical analysis, we selected sample data including research articles published in Science and Nature, top 1% highly cited papers, and zero-cited papers for the years 2005, 2010, 2015, 2020, and 2025. We calculated the Knowledge Centrifugality scores for these papers and examined their potential influencing factors. The results indicate that team size exerts a significant negative effect on paper novelty ( β =-0.005, p<0.001), meaning larger team size is less conducive to enhancing the novelty of research outputs. Conversely, the number of references shows a significant positive correlation with paper novelty ( β

=0.027,p<0.001), which means that a greater number of references is associated with a moderate improvement in a paper's novelty. The proposed Knowledge Eccentricity offers strong timeliness and operability, allowing for the evaluation of a paper's novelty immediately upon its publication.

## 1 引言

对研究成果新颖性进行评价，筛选潜在变革性研究，加速前瞻培育和布局，提升基础研究的原始创新能力，是科研管理部门长久以来的夙愿和挑战。科学发展离不开对已有知识的重新组合，而新颖的研究通常依赖对现有知识的"非常规重组"[1-4]，涉及跨越传统学科边界，将原本互不关联的知识元、方法、理论或数据集进行创造性整合[5-8]。例如，Uzzi 等认为新颖性是对已有知识的非常规重组[1]，Wang 等认为新颖性是已有知识"意料之外"的组合[9]，Foster 等认为新颖性包括涉足人类尚未意识到的知识、首次将不同学科知识进行组合、将学科内尚未联系的知识进行组合三种类型[10]。

从"非常规重组"视角，对学术成果新颖性进行测度的方法通常基于期刊、关键词、引文等。如 Uzzi's z-score 指标及其衍生指标[1, 2, 9-11]、结构洞[12, 13]，颠覆性指数[14-16]、新词占比等[7]。Boudreau 等通过分析在项目申报书中出现但未收录于 PubMed 数据库中 MeSH 词的比例来测度项目申报书的新颖性，发现评审专家对新颖性较高的申报书评价较负面[17]。Wu 等基于引文网络，提出现有科研成果对已有成果替代程度的指标（即颠覆性指数），认为当后续研究更多引用待评价成果且较少引用待评价成果的参考文献时，则待评价成果的新颖性较强[14]。近两年，随着深度学习技术的进步，基于语言大模型进行研究成果新颖性测度的方法也随之出现[18-20]。

尽管上述方法或指标在一定程度上反映了研究成果的新颖性，但在"及时性"和"可操作性"方面仍存在显著局限：一是，及时评价的困境：基于引用数据的评价指标需依赖时间的积累，难以对新发表研究成果的新颖性作出即时判断，如颠覆性指数及其衍生指标[21]。而且一旦引入待评价成果的施引文献信息，则该指标就易于被操纵。二是，可操作性不强。被引次数、期刊影响因子、期刊分区等指标尽管存在较多弊端，但长期被用于科研评价的原因之一是其操作简单，而 Uzzi's z-score 指标及其衍生指标、结构洞、新词占比等尽管及时性较强，但因其计算复杂度较高，很难普及。

本研究基于论文的参考文献信息，从已有知识"非常规重组"视角，提出一种衡量论文新颖性的指标：即知识离心度（Knowledge Eccentricity, KE），该指标在论文发表时即可对新颖性进行测度。为提高 KE 指标的可操作性，本研究开发了基于 DeepSeek 的 KE 指标计算智能体和批量计算平台，科研管理部门仅需提供待评价成果的 doi，便可迅速计算该成果的新颖性，且计算过程直观、透明。

## 2 新颖性指标设计与研究思路

### 2.1 指标设计

本研究认为，待评价论文所引用的参考文献之间集聚程度较强，能够反映待评价论文的知识离心度较弱，新颖性不强。即如果一篇学术论文所引用的都是密集交织在一起的参考文献，那么该篇学术论文中的知识并未偏离原有的知识基础，而是依赖已有知识的常规组合。反之，则属于新颖性较强的成果，具有潜在的颠覆性。为测度学术论文 v 的知识离心度（Knowledge Eccentricity，$KE_v$），我们提出以下公式：

$$KE_v = 1 - \sqrt[3]{\frac{2L}{N*(N-1)}}$$

　　其中，L 表示论文 v 的参考文献之间存在引用关系的数量，N 表示论文 v 的参考文献数量。图 1 展示了三种不同类型的知识组合模式，其中橘色节点 v 表示待评价论文，蓝色节点表示待评价论文 v 的参考文献。具体计算步骤为：第一步：获取待评价论文的参考文献集合 R={Ref$_1$, Ref$_2$,…,Ref$_n$}。第二步：获取集合 R 中每一篇参考文献的参考文献集合 RR={RRef$_1$, RRef$_2$,…,RRef$_n$}。第三步：计算集合 R 中参考文献配对在集合 RR 中出现与否（记做 $l_v$），出现记为 1，反之为 0。第四步：将全部 $l_v$ 累加，L=$\sum_n l_v$，最终根据公式计算 $KE_v$，概念模型见图 1*。

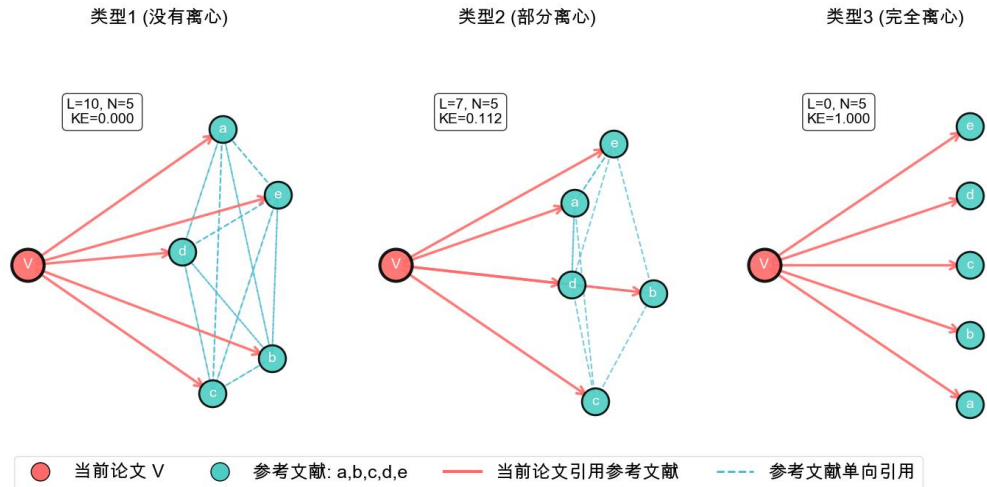

图 1　三种不同类型的知识组合情况。Type I 中，全部参考文献均存在引用行为($KE_v$=0)；Type II 中有部分参考文献中存在引用行为($KE_v$=0.112)；Type III 中参考文献之间不存在引用行为($KE_v$=1)。

## 1.2 研究思路

　　Science 和 Nature 两本期刊在学术界具有较高的名望，成果发表在上述两本期刊是一种荣誉，有助于带来更多的关注和引文影响力。根据引文规范理论，高被引论文是存在较强引文影响力的论文，而零被引论文则是长期以来未被科学界认可的成果。基于此，本研究收集 2005 年、2010 年、2015 年、2020 年和 2025 年发表在 Science 和 Nature 上的研究型论文、Top1%高被引论文和零被引论文作为样本数据，计算上述论文的 KE 值，并对其潜在的影响因素进行分析，研究思路见图 2。

---

* KE 指标在线计算：https://www.coze.cn/space/7446604282029588489/bot/7581092641727758345

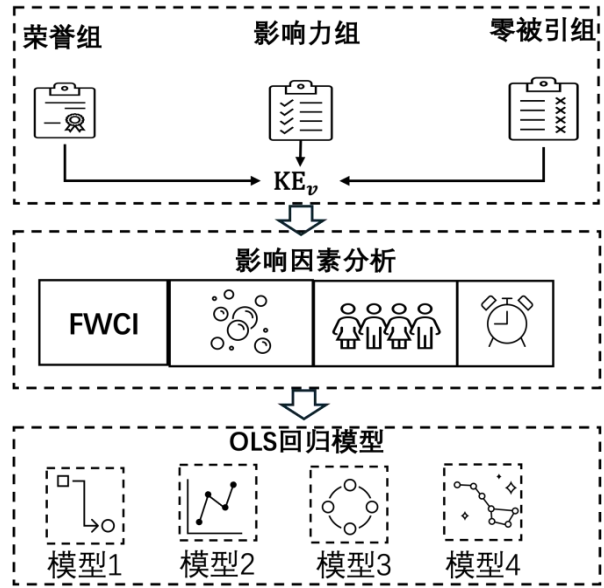

图 2　研究思路

## 2　研究结果

本研究的数据来源于 OpenAlex 数据库，它涵盖了广泛的学术出版物、作者、机构和引用关系。在删除参考文献为 0 的论文后，三组论文合计 20184 篇。下面分别从发表时间、研究领域、团队规模等对新颖性指标的影响进行分析。

### 2.1　时间与分组对论文新颖性的影响

结果显示，不同时间发表的论文，其新颖性指标整体上介于 0.530-0.602 之间，未产生较大波动，但存在轻微时间变化趋势，即发表越早的论文整体的新颖性偏低。组间比较结果显示，荣誉组和影响力组（$t=-22.724$, $p<0.001$）、荣誉组和零被引组（$t=-20.611$, $p<0.001$）、影响力组和零被引组（$t=-8.121$, $p<0.001$）论文的新颖性存在统计学差异，整体呈现出"零被引组>影响力组>荣誉组"的趋势。

KE 值分布的多样性可以从一定程度上反映不同组内论文新颖性程度是否一致。由于香农指数是衡量系统多样性的常用指标[22]，因此，本研究采用香农指数对各组论文 KE 值分布的多样性进行分析，发现高影响力组的多样性最高（香农指数为 1.278），其次是零被引组（香农指数为 1.214）和荣誉组（香农指数为 1.099），与采用辛普森指数和基尼-辛普森指数计算结果的趋势一致（见附录 A），提示高影响力论文的新颖性变动较大，其次是零被引论文，而发表论文发表在 Science 或 Nature 上的新颖性变动较小。

以三组论文新颖性的均值 0.581 为阈值，荣誉组中 KE≥0.581 的论文占比最少（35.2%），其次是影响力组（51.5%）和零被引组（56.8%）。另外，荣誉组和影响力组论文的新颖性分布与零被引组明显不同，零被引组论文的新颖性呈现明显的极化趋势，即 KE=0 和 1 的论文较另外两组多，特别是 KE=1 的论文数量，提示新颖性过高的研究在短时间通常不被认可，见图 3。

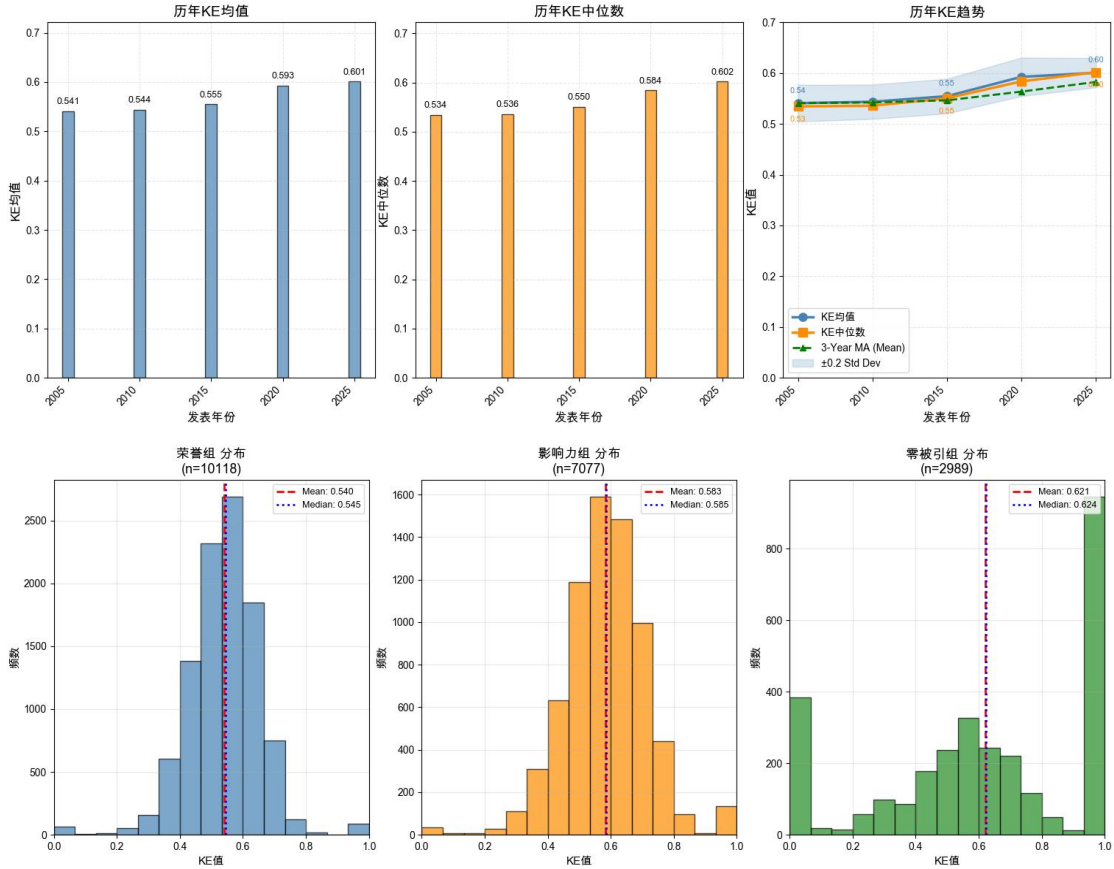

图 3 KE 值在不同时间和组间的分布图

## 2.2 研究领域对论文新颖性的影响

在进行研究领域大类统计后发现，数据源于社会科学（Social Sciences）、健康科学（Health Sciences）、物理科学（Physical Sciences）和生命科学（Life Sciences）四大领域，荣誉组、影响力组和零被引组各领域构成等详细信息见附录 B。

进行方差齐性检验后，发现不同研究领域具有可比性，因此采用 One-way ANOVA 进行多组比较。结果显示不同研究领域论文的新颖性存在显著的统计学差异($p<0.001$)，见表 1。为进一步分析各组间差异，采用 Tukey HSD 检验进行两两比较，结果显示社会科学领域论文的新颖性显著高于其他领域（$p<0.001$），健康科学和物理科学两个领域论文的新颖性差异没有显著差异($p>0.05$)，生命科学较健康科学、物理科学两个领域论文的新颖性低($p<0.001$)，见表 2。

表 1 不同学科论文新颖性的基本情况

| 研究领域 | 样本量 | 均值 | 标准差 | 中位数 | 下四分位 | 上四分位 |
|---|---|---|---|---|---|---|
| 物理科学 | 8509 | 0.561 | 0.157 | 0.558 | 0.482 | 0.637 |
| 生命科学 | 5101 | 0.538 | 0.118 | 0.544 | 0.471 | 0.610 |
| 健康科学 | 3511 | 0.555 | 0.161 | 0.557 | 0.478 | 0.638 |
| 社会科学 | 2903 | 0.649 | 0.264 | 0.646 | 0.545 | 0.793 |

表 2 不同领域论文新颖性的 Tukey HSD 检验

| 组 1 | 组 2 | 均差* | 校正 p | 下限 | 上限 | 拒绝原假设 |
|---|---|---|---|---|---|---|
| 健康科学 | 生命科学 | -0.018 | 0.000 | -0.027 | -0.008 | TRUE |

| | | | | | |
|---|---|---|---|---|---|
| 健康科学 | 物理科学 | 0.006 | 0.322 | -0.003 | 0.015 | FALSE |
| 健康科学 | 社会科学 | 0.094 | 0.000 | 0.083 | 0.105 | TRUE |
| 生命科学 | 物理科学 | 0.024 | 0.000 | 0.016 | 0.031 | TRUE |
| 生命科学 | 社会科学 | 0.112 | 0.000 | 0.102 | 0.122 | TRUE |
| 物理科学 | 社会科学 | 0.088 | 0.000 | 0.079 | 0.098 | TRUE |

*均差为组 2 的均值减组 1 的均值。

## 2.3 团队规模、参考文献量和 FWCI 对论文新颖性的影响

科研团队是为实现特定科学目标，由具备不同知识、技能和专长的科研人员组成的协作群体。随着科学活动专业程度日趋提升以及跨学科解决问题的需求增加，大科研团队比例逐渐增多，发表在高期刊分区上的论文尤为明显[23, 24]。领域权重引用影响力（FWCI）是文献实际被引用次数，与该领域预期引用总量平均值的比值，旨在领域间论文影响力的横向比较[25]。当 FWCI>1 时，表示论文被引次数超过全球预期引用量的平均水平。

将团队规模、参考文献量和 FWCI 对论文新颖性的影响进行分析，并按照四分位数法，进一步将团队规模分为小团队（$Q_1$，$P_{25}$）、中小团队（$Q_2$，$P_{50}$）、中大团队（$Q_3$，$P_{75}$）和大团队（$Q_4$）四组，将参考文献量划分为少量（$Q_1$，$P_{25}$）、较少（$Q_2$，$P_{50}$）、较多（$Q_3$，$P_{75}$）和多参考文献（$Q_4$）四组。由于 FWCI 存在等于 0 的情况，为确保组内样本量均衡，将 FWCI 进一步分为零、低、中低、中高和高 FWCI 组。剔除没有作者、FWCI 缺失等字段，最终对 20158 篇论文的团队规模、FWCI 与对论文新颖性影响进行分析，对 20149 篇论文的参考文献量与论文新颖性影响进行分析。

结果显示，团队规模与论文新颖性呈弱负相关 (r=-0.023, p<0.001)，参考文献量与论文新颖性呈弱正相关 (r=0.160, p<0.001)，FWCI 与论文新颖性呈弱正相关 (r=0.051, p<0.001)，且各变量内的分组间存在显著的统计学差异（p<0.001）。上述结果提示：小团队和大团队、少参考文献和多参考文献、零 FWCI 和高 FWCI 的论文通常具有较高的新颖性，见图 4。

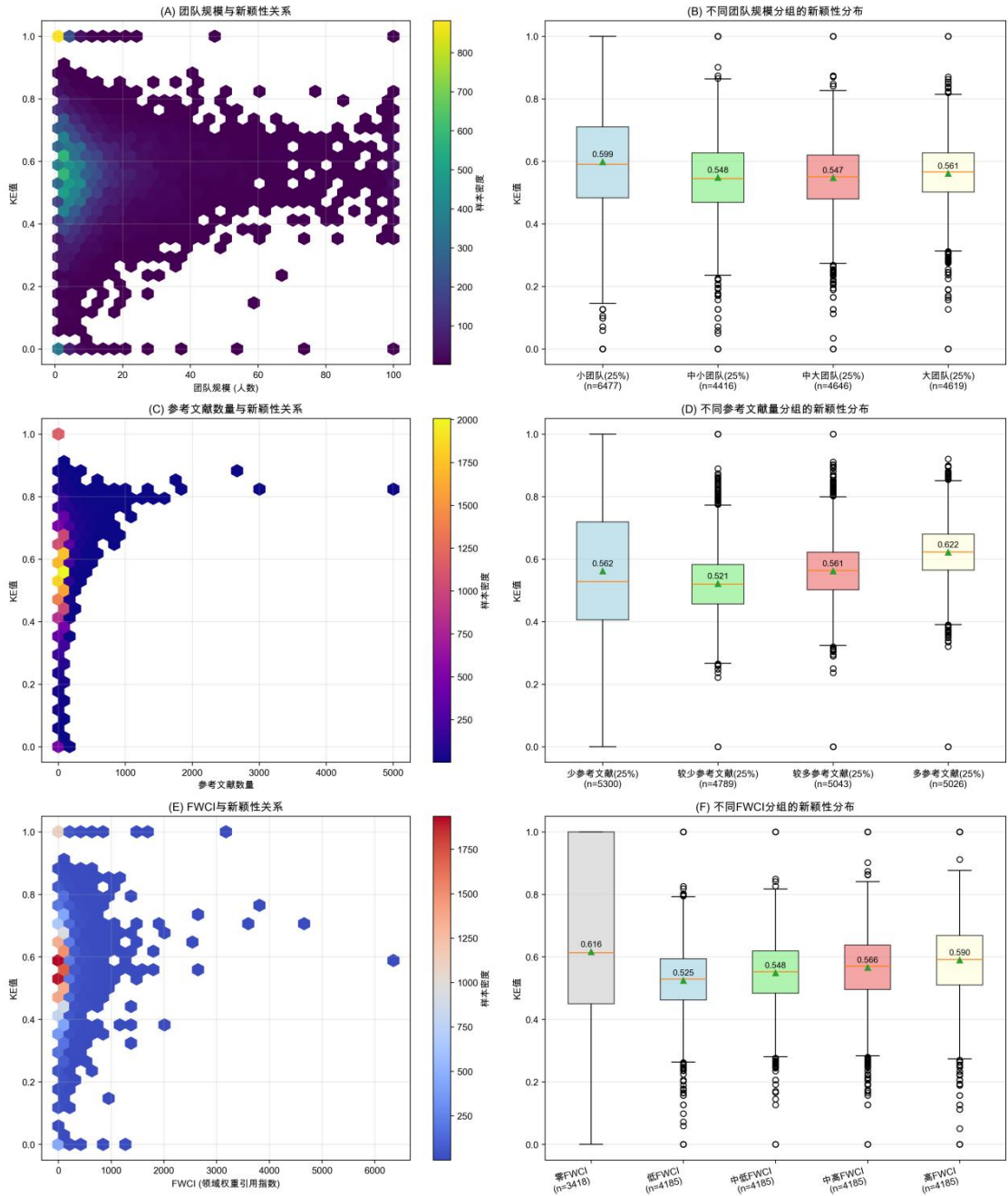

图 4 团队规模、参考文献量和 FWCI 对论文新颖性的影响

## 2.4 论文新颖性影响因素的 OLS 回归分析

将论文被引次数、参考文献量、FWCI、团队规模作为自变量，论文新颖性作为因变量，发表年份和研究领域作为控制变量，进行标准化 OLS 回归分析。对自变量进行标准化和一致性处理，并采用方差膨胀因子（VIF）进行多重共线性检验，结果显示 VIF 介于 1-1.5 之间，提示上述自变量的共线性较轻。回归模型分析结果显示，被引次数对 KE 具有微弱但显著的正向影响（$\beta=0.004$, $p<0.05$），提示学术认可度可能促进知识创新（模型 1）。模型 2 引入参考文献量后，其表现出极强的正向效应（$\beta=0.027$, $p<0.001$），而此前被引次数的影响不再显著，提示参考文献量对论文新颖性的解释力较强。模型 3 加入 FWCI 值，但该变量未呈现显著影响，提示学科间相对影响力的作用有限。模型 4 进一步纳入团队规模，发现其具有显著的负向效应（$\beta=-0.005$, $p<0.001$），提示团队规模越大，越不利于成果的新颖性，而参考文献量的增加对论文新颖性具

有正向作用,见表 3。

从模型整体拟合度看,调整 R²从 0.061 逐步提升至 0.086,提示自变量对论文新颖性具有一定解释力。其中参考文献量的稳健显著性尤为突出,强调了知识基础对创新的关键作用,为后续创新研究提供了理论依据。上述结果与本研究随机森林特征重要性排序结果一致(见附录 C)。

表 3　论文新颖性影响因素的 OLS 回归分析

|  | 模型 1 | 模型 2 | 模型 3 | 模型 4 |
| --- | --- | --- | --- | --- |
| 被引次数 | 0.004** (0.002) | 0.001 (0.321) | 0.002 (0.195) | 0.002 (0.179) |
| 参考文献量 |  | 0.027*** (0.000) | 0.028*** (0.000) | 0.028*** (0.000) |
| FWCI 值 |  |  | -0.001 (0.382) | -0.001 (0.576) |
| 团队规模 |  |  |  | -0.005*** (0.000) |
| N | 20184 | 20184 | 20184 | 20184 |
| $R^2$ | 0.062 | 0.086 | 0.086 | 0.087 |
| 调整 $R^2$ | 0.061 | 0.086 | 0.086 | 0.086 |

*控制变量为研究领域、发表年份, (*即 p< 0.05, **即 p< 0.01, ***即 p< 0.001)

## 3. 结论与讨论

传统科研评价过度依赖论文数量、被引次数、影响因子等指标,在衡量学术成果新颖性方面存在不足。科学创新的核心机制之一在于对现有知识进行新颖且非常规的组合,尽管学术界普遍存在依赖成熟知识的路径倾向,但突破性创新往往源于对不熟悉主题的探索及跨领域的知识整合[10, 21]。然而,高度新颖性的研究伴随学科交叉性强、知识结构"非常规"特征,往往挑战现有范式,因而在传统评价体系中难以获得及时认可。例如,辛顿(Geoffrey E. Hinton)关于神经网络的早期研究、麦克林托克(Barbara McClintock)的"跳跃基因"假说、安布罗斯(Victor Ambros)关于微小核糖核酸(miRNA)及基因调控中机制方面的探索等。

从已有知识"非常规重组"视角出发,本研究认为,当论文所引用的参考文献紧密聚集在一起时,其知识基础往往较为成熟,则该论文整合跨学科知识的能力有所欠缺。基于此,提出测度知识非常规重组的知识离心度指标,并对其影响因素进行分析。研究认为,当不考虑学科领域和发表年份时,零被引论文的新颖性从整体上高于具有较高引文影响力的论文,而发表在 Science 和 Nature 上的论文其新颖性则普遍不如零被引论文和高影响力论文。从各组论文新颖性指标的多样性分布看,零被引组论文的新颖性波动(香农指数为 1.214)相对较大,一方面与零被引组中社会科学领域的论文占比(42.8%)较荣誉组(3.9%)和影响力组(17.8%)有关,另一方面,从各组新颖性等于 1 的论文数量看,零被引组新颖性等于 1 的论文数量远远高于其他两组,从一定程度上印证了高度新颖的研究,往往不能得到及时认可的情形,呈现出"睡美人"特征。另外,高影响力论文的新颖性变动最大,而发表在 Science 或 Nature 上的论文新颖性变动最小,提示研究成果的引文影响力与新颖性并不完全一致,而发表在 Science 或 Nature 上的论文因其异常严格的同行评审和发表压力,往往在知识重组过程中保持了必要的张力[26]。

从研究领域对论文新颖性的影响来看,不同研究领域论文的新颖性不存在可比性,例如社会科学领域的论文新颖性普遍高于自然科学领域的论文,提示在采用 KE 值进行论文新颖性测度和领域间横向比较时,要控制研究领域的影响。另外,研究认为小团队研究成果的新颖性通常最强,这与 Wu 等[14]人的研究结论基本一致。 FWCI 与论文新颖性呈弱正相关,反映被引次数与论文新颖性指标存在一定的正相关关系。

控制研究领域和发表年份后,回归结果发现团队规模和参考文献量是影响论文新颖性的主要因素,团队规模对论文新颖性的影响是负向的,即团队规模越大,越不利于论文的新颖性提升;而参考文献数量与论文

新颖性呈现正向关系，即当论文的参考文献数量越多时，论文的新颖性也存在一定程度的提升，提示新颖性的研究多建立在大量已有知识基础之上。

综上所述，本研究提出一种论文新颖性测度的新指标，并对其影响因素进行了分析，认为该指标可及时、方便地对论文新颖性进行评价，为零被引论文新颖性的识别提供了新思路。研究成果新颖性受团队规模和已有知识基础的影响，与引文及其相关衍生指标无关，如被引次数，FWCI 等。值的注意的是，知识"非常规重组"只是知识创新的一种典型方式，它既不完全是知识创新的充分条件，也不完全是必要条件，即并非所有知识的"非常规重组"都能产生新颖的研究，而新颖的研究也不全是知识"非常规重组"的结果。

## 参 考 文 献

[1] UZZI B, MUKHERJEE S, STRINGER M, et al. Atypical combinations and scientific impact [J]. Science, 2013, 342(6157): 468-72.
[2] LEE Y-N, WALSH J P, WANG J. Creativity in scientific teams: Unpacking novelty and impact [J]. Research Policy, 2015, 44(3): 684-97.
[3] SHI F, EVANS J. Surprising combinations of research contents and contexts are related to impact and emerge with scientific outsiders from distant disciplines [J]. Nat Commun, 2023, 14(1): 1641.
[4] SCHILLEBEECKX S J D, LIN Y, GEORGE G, et al. Knowledge Recombination and Inventor Networks: The Asymmetric Effects of Embeddedness on Knowledge Reuse and Impact [J]. Journal of Management, 2020, 47(4): 838-66.
[5] SHIBAYAMA S, WU Z, YIN D, et al. State of the Art of Novelty Indicators [J]. SSRN, 2025.
[6] ZHAO Y, ZHANG C. A review on the novelty measurements of academic papers [J]. Scientometrics, 2025, 130(2): 727-53.
[7] BORNMANN L, TEKLES A, ZHANG H H, et al. Do we measure novelty when we analyze unusual combinations of cited references? A validation study of bibliometric novelty indicators based on F1000Prime data [J]. Journal of Informetrics, 2019, 13(4).
[8] WAGNER C S, WHETSELL T A, MUKHERJEE S. International research collaboration: Novelty, conventionality, and atypicality in knowledge recombination [J]. Research Policy, 2019, 48(5): 1260-70.
[9] WANG J, VEUGELERS R, STEPHAN P. Bias against novelty in science: A cautionary tale for users of bibliometric indicators [J]. Research Policy, 2017, 46(8): 1416-36.
[10] FOSTER J G, RZHETSKY A, EVANS J A. Tradition and Innovation in Scientists' Research Strategies [J]. American Sociological Review, 2015, 80(5): 875-908.
[11] LUO Z, LU W, HE J, et al. Combination of research questions and methods: A new measurement of scientific novelty [J]. Journal of Informetrics, 2022, 16(2).
[12] CHEN C, DUBIN R, KIM M C. Emerging trends and new developments in regenerative medicine: a scientometric update (2000 - 2014) [J]. Expert Opin Biol Ther, 2014, 14(9): 1295-317.
[13] RZHETSKY A, FOSTER J G, FOSTER I T, et al. Choosing experiments to accelerate collective discovery [J]. Proc Natl Acad Sci U S A, 2015, 112(47): 14569-74.
[14] WU L, WANG D, EVANS J A. Large teams develop and small teams disrupt science and technology [J]. Nature, 2019, 566(7744): 378-82.
[15] LEAHEY E, LEE J, FUNK R J. What Types of Novelty Are Most Disruptive? [J]. American Sociological Review, 2023, 88(3): 562-97.
[16] LIN Y, EVANS J A, WU L. New directions in science emerge from disconnection and discord [J]. Journal of Informetrics, 2022, 16(1).
[17] BOUDREAU K J, GUINAN E C, LAKHANI K R, et al. Looking Across and Looking Beyond the Knowledge Frontier: Intellectual Distance, Novelty, and Resource Allocation in Science [J]. Manage Sci, 2016, 62(10): 2765-83.
[18] WANG Z, WANG Z, ZHANG G, et al. A hybrid graph and LLM approach for measuring scientific novelty via knowledge recombination and propagation [J]. Expert Systems with Applications, 2026, 298.
[19] HUANG S, LU W, XU Z, et al. Identifying potentially disruptive research via a comparative power-based large model [J]. Information Processing & Management, 2025, 62(6).
[20] WANG Z, ZHANG H, CHEN J, et al. An effective framework for measuring the novelty of scientific articles through integrated topic modeling and cloud model [J]. Journal of Informetrics, 2024, 18(4).
[21] 杨杰，邓三鸿，王昊. 科学研究的颠覆性创新测度——相对颠覆性指数 [J]. 情报学报, 2023, 42(9): 1052-64.
[22] SHANNON C E. A Mathematical Theory of Communication [J]. Bell System Technical Journal, 1948, 27(3): 379-423.
[23] TANG X, SHI W, WU R, et al. The expansion of team size in library and information science (LIS): Is bigger always better? [J]. Journal of Information Science, 2023.
[24] LIN Y, LI L, WU L. Team size and its negative impact on the disruption index [J]. Journal of Informetrics, 2025, 19(3).
[25] SCELLES N, TEIXEIRA DA SILVA J A. Making the impact of publications within a field comparable by improving the field-weighted citation impact (FWCI): the case of sport management [J]. Scientometrics, 2025, 130(3): 1571-86.

[26] HOYNINGEN-HUENE P, LOHSE S. On naturalizing Kuhn's essential tension [J]. Studies in History and Philosophy of Science Part A, 2011, 42(1): 215-8.

# 附录

**附录 A**

表 1 三组论文的多样性分布

| 组别 | 样本量 | 唯一领域数 | 香农指数 | 辛普森指数 | 基尼-辛普森指数 |
|---|---|---|---|---|---|
| 荣誉组 | 10008 | 4 | 1.099 | 0.625 | 0.625 |
| 影响力组 | 7077 | 4 | 1.278 | 0.694 | 0.694 |
| 零被引组 | 2939 | 4 | 1.214 | 0.673 | 0.673 |

**附录 B**

表 2 不同年份和领域论文量

| publication_year | Physical Sciences | Life Sciences | Health Sciences | Social Sciences |
|---|---|---|---|---|
| 2005 | 1667 | 1120 | 644 | 640 |
| 2010 | 1553 | 1113 | 595 | 552 |
| 2015 | 1695 | 1072 | 701 | 501 |
| 2020 | 1501 | 900 | 1089 | 852 |
| 2025 | 2093 | 896 | 482 | 358 |

| | 研究领域 | 文献数量 | 组内占比 (%) | 总占比 (%) |
|---|---|---|---|---|
| 影响力组 | Physical Sciences | 3094 | 43.72 | 15.45 |
| | Health Sciences | 1848 | 26.11 | 9.23 |
| | Social Sciences | 1258 | 17.78 | 6.28 |
| | Life Sciences | 877 | 12.39 | 4.38 |

| | 研究领域 | 文献数量 | 组内占比 (%) | 总占比 (%) |
|---|---|---|---|---|
| 荣誉组 | Physical Sciences | 4453 | 44.49 | 22.24 |
| | Life Sciences | 4037 | 40.34 | 20.16 |
| | Health Sciences | 1130 | 11.29 | 5.64 |
| | Social Sciences | 388 | 3.88 | 1.94 |

| | 研究领域 | 文献数量 | 组内占比 (%) | 总占比 (%) |
|---|---|---|---|---|
| 零被引组 | Social Sciences | 1257 | 42.77 | 6.28 |
| | Physical Sciences | 962 | 32.73 | 4.8 |
| | Health Sciences | 533 | 18.14 | 2.66 |
| | Life Sciences | 187 | 6.36 | 0.93 |

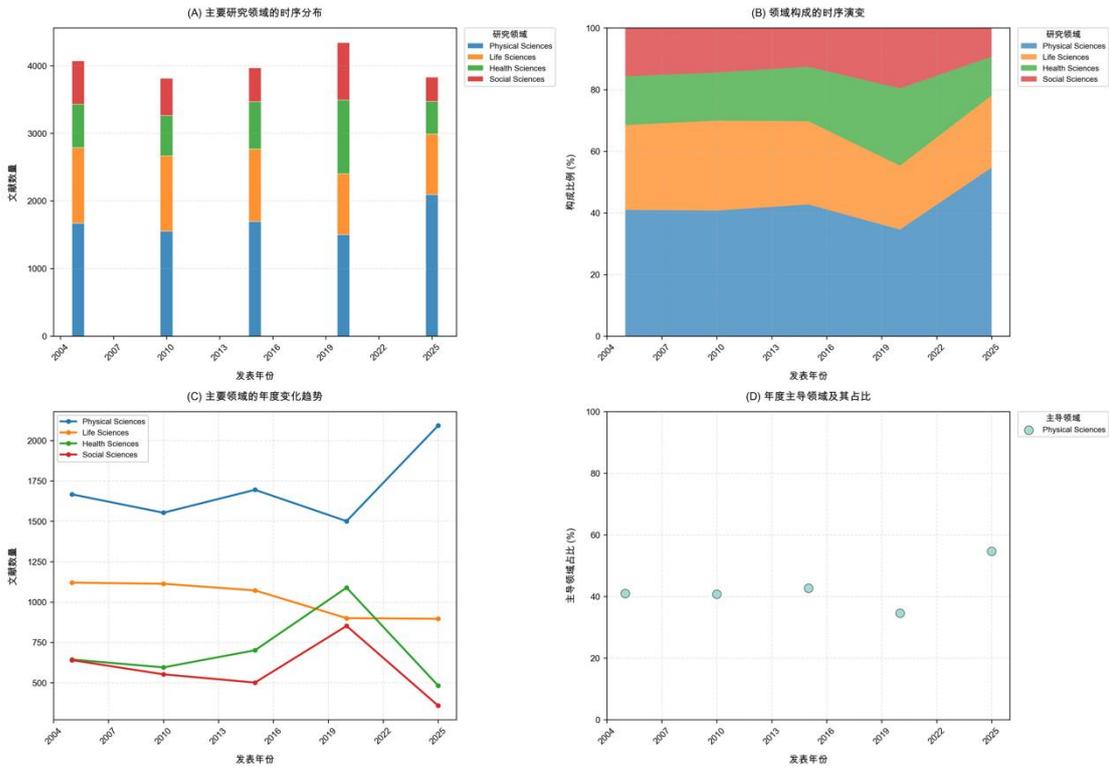

图 1　不同年份和领域的文献量

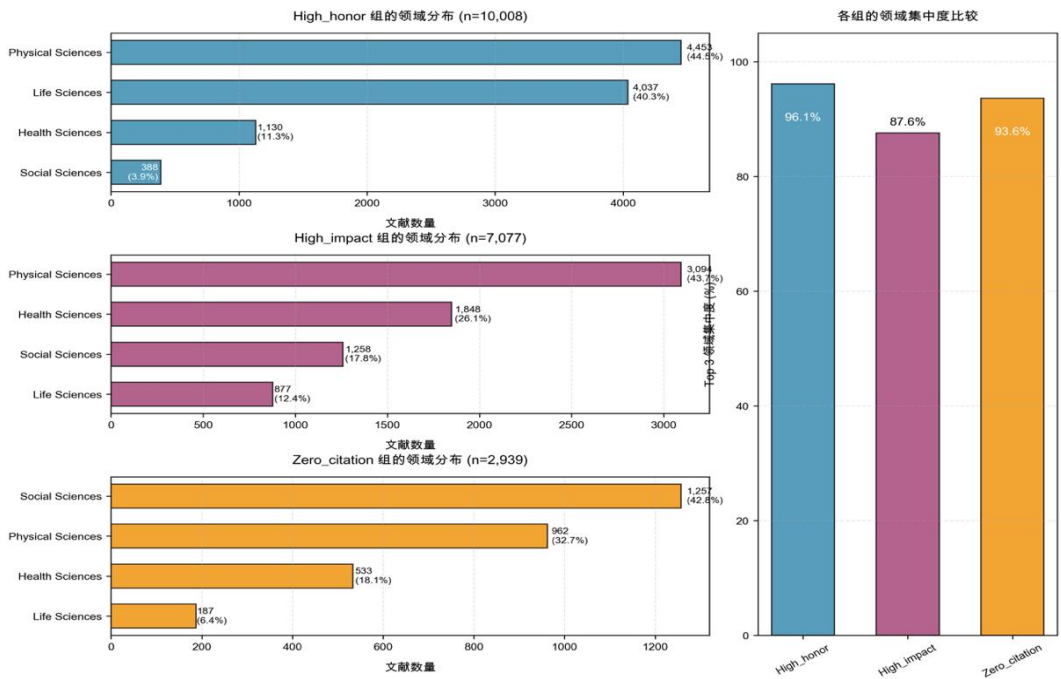

图 2　不同年份、组别、研究领域的文献量

附录 C

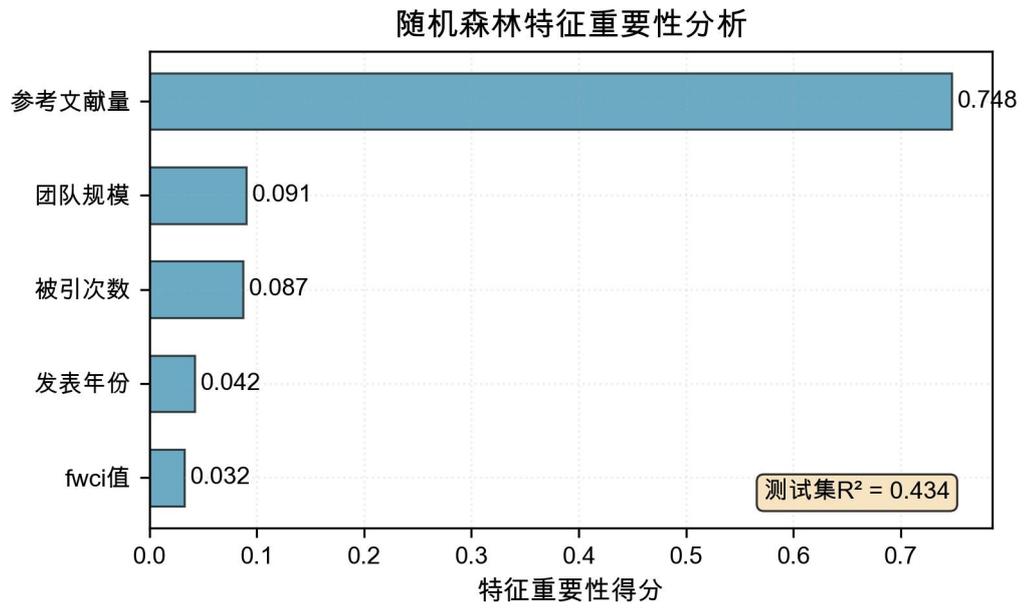

图 3　随机森林特征重要性得分